\documentclass[english,aps,pre, preprint, onecolumn]{revtex4}
\usepackage[T1]{fontenc}
\usepackage[latin9]{inputenc}
\usepackage{float}
\usepackage{amsmath}
\usepackage{graphicx}
\usepackage[caption=false]{subfig}
\makeatletter
\@ifundefined{textcolor}{}
{%
 \definecolor{BLACK}{gray}{0}
 \definecolor{WHITE}{gray}{1}
 \definecolor{RED}{rgb}{1,0,0}
 \definecolor{GREEN}{rgb}{0,1,0}
 \definecolor{BLUE}{rgb}{0,0,1}
 \definecolor{CYAN}{cmyk}{1,0,0,0}
 \definecolor{MAGENTA}{cmyk}{0,1,0,0}
 \definecolor{YELLOW}{cmyk}{0,0,1,0}
}

\makeatother

\usepackage{babel}
\begin{document}

\preprint{\%This line only printed with preprint option}

\title{A quantitative analysis of the emergence of memory in the viscously coupled dynamics of colloids}

\author{Shuvojit Paul}

\author{Randhir Kumar}

\author{Ayan Banerjee}
\email{ayan@iiserkol.ac.in}

\affiliation{Dept of Physical Sciences, Indian Institute of Science Education
and Research, Mohanpur, Kolkata, India 741235}
\begin{abstract}
We provide a quantitative description of the evolution of memory from the apparently random Markovian dynamics of a pair of optically trapped colloidal microparticles in water. The particles are trapped in very close proximity of each other so that the resultant hydrodynamic interactions lead to non-Markovian signatures manifested by the double exponential auto-correlation function for the Brownian motion of each particle. In connection with the emergence of memory in this system, we quantify the storage of energy and demonstrate that a pair of Markovian particles - confined in individual optical traps in a viscous fluid - can be described in the framework of a single Brownian particle in a viscoelastic medium. We define and quantify the equivalent storage and loss moduli of the two-particle system, and show experimentally that the memory effects reduce with increasing particle separation and increase with a skewed stiffness ratio between the traps.
\end{abstract}
\maketitle


The dynamics of isolated colloidal particles in a viscous fluid at
low Reynold's number, and with large viscous drag force which dominates
inertial effects, is memory-less and therefore Markovian - being solely
determined by instantaneous forces \cite{purcell1977purcell}.
At very short time scales, however, inertia manifests itself so that
particle motion becomes correlated \cite{Einstein05}. More complex
memory effects arise due to the motion of the particle in the fluid
due to the evolution of ``hydrodynamic memory'' \cite{balder70}
that is observed at the so-called vorticity diffusion time scales,
and has been experimentally observed only recently due to the extremely
high spatial and temporal resolution involved \cite{huang2011}. In
diffusive time scales that are relatively simple to obtain in experiments,
memory effects are easily observed in the Brownian motion of colloidal
particles in complex fluids, or in living matter and active particles.
Interestingly, the coupled dynamics of colloidal particles show intriguing
effects even in viscous media. Thus, recent work has pointed out
that interacting (and therefore out of thermal equilibrium) active
particles with correlated noises in a viscous fluid can be interpreted
to be in equilibrium with a viscoelastic bath \cite{fodor2016}. In
addition, it has also been experimentally demonstrated that
coupled over-damped optically trapped Brownian harmonic oscillators
in a viscous fluid show non-zero cross-correlation \cite{meiners1999}, which indicates the presence of memory in such systems.

Recent theoretical work \cite{dua2011non} with a view to explain
experimentally measured protein dynamics \cite{yang2003}, has demonstrated
that non-Markovian signatures may, in general, evolve from a superposition
of distinct and uncorrelated Markovian processes. This is because,
a simple superposition of two such processes may lead to a double
exponential auto-correlation function, which obviously signifies a
non-Markovian process since Doob's theorem stipulates that the auto-correlation
of a Gaussian, stationary, and Markovian process necessarily needs
to be a single exponential. To the best of our knowledge, direct experimental
verification of this statement is not available presently. Single
colloidal particles confined in optical tweezers in a viscous medium
act as an ideal test bed for this, since the dynamics of such particles
is strictly Markovian in the diffusive regime. However, when two such
particles are trapped in close proximity so that they are coupled
hydrodynamically, interesting and unexpected phenomena appear. Our
recent work in this system has revealed a frequency maximum in the
amplitude of the mutual response of this system to an external excitation
\cite{paul2017}. A natural question would thus be whether this coupled
system would display stationary, but non-Markovian signatures with
the individual Markovian responses being superposed owing to the hydrodynamic
coupling. The evolution of memory can then be studied very carefully,
and the system be cast in the framework of one of the simplest non-Markovian
processes accessible by optical tweezers - namely the dynamics of
a single trapped colloidal particle in a viscoelastic fluid.

In this paper, we provide theoretical and experimental confirmation of this supposition. We work with a pair of Brownian particles trapped separately in dual optical tweezers. Theoretically, we observe that the auto-correlation functions of the individual particles in this system are double exponential in nature, which confirms that the statistics of the Brownian fluctuation of
each particle is non-Markovian. Thereafter, we demonstrate that this
system can be directly mapped to a single Brownian particle trapped
in a linear viscoelastic medium with a single polymer relaxation time
- the double exponential nature of the auto-correlation function of
the latter arising due to the trap time constant and the elasticity
of the medium. In contrast, for the particles in the dual trap in
the viscous medium, the time constants are due to the two traps, which
are reflected in the motion of each particle by the entrained fluid
which couples the particles with each other. We also quantify
the storage modulus $G'$ and loss modulus $G''$ for this system
that reflect the storage and loss of energy in the fluid \cite{mason2000estimating,tassieri2010measuring}.
The storage of energy directly reflects the emergence of memory in
this system, and provides a complete understanding of the underlying
interesting physics of this problem. We proceed to first layout the
theoretical formalism for this system.

\emph{Theory.} We consider two hydrodynamically coupled Brownian spherical
particles executing Brownian motions in their respective optical traps.
The equation of motion describing the stochastic trajectories of these
two particles can be represented in matrix form, as \cite{paul2017direct,paul2018two,gardiner1985handbook}
\begin{equation}
m_{i}\dot{\boldsymbol{v}}_{i}=-\boldsymbol{\gamma}_{ij}\boldsymbol{v}_{j}-{\bf \boldsymbol{\nabla}}_{i}U+{\bf \boldsymbol{\xi}}_{i}\label{eq:1}
\end{equation}
\begin{equation}
{\bf \dot{r}}_{i}=\boldsymbol{v}_{i}\label{eq:2}
\end{equation}
where, $i,\,j=1,\,2$ are particle identifiers, with ${\bf r}_{i}$ and $\boldsymbol{v}_{i}$
being the position and velocity, respectively, of the i-th particle of
mass $m_{i}$, $\mathrm{U}$ is the potential created by the optical
trap, and $\boldsymbol{\xi}_{i}$ is the zero-mean, Gaussian distributed
thermal stochastic noise related to the i-th particle. The variance
of the noise is determined by the fluctuation-dissipation relation
$\left\langle \boldsymbol{\xi}_{i}(t)\boldsymbol{\xi}_{j}(t')\right\rangle =2k_{B}T\boldsymbol{\gamma}_{ij}\delta(t-t')$
which implies instantaneous correlation. $\boldsymbol{\gamma}_{ij}$
are the second-rank friction tensors encoding the velocity-dependent
dissipative forces mediated by the fluid. The optical potential is
given by $U(t)=\frac{1}{2}\sum k_{i}({\bf r}_{i}-{\bf r}_{i}^{0})^{2}$
with ${\bf r}_{i}^{0}$ being the position of the potential minimum of
the i-th optical trap and $k_{i}$ is the corresponding stiffness.
A variety of methods can be used to calculate the friction tensors from Stokes equations in the limit of slow viscous flow in the fluid
\cite{ladd1988hydrodynamic,mazur1982many}. The assumption of slow
viscous flow is valid $\omega\tau_{\nu}<<1$, where $\tau_{\nu}=\rho L^{2}/\eta$
is the vorticity diffusion time scale. $\rho$, $L$, $\eta$ are
the density, length scale and viscosity of the medium respectively.
$\tau_{\nu}$ is around $<<2.7\,\mu s$ in our case. To leading order
\begin{equation}
\boldsymbol{\gamma}_{ij}(\boldsymbol{r}_{i},\boldsymbol{r}_{j})=\delta_{ij}\boldsymbol{I}\gamma_{i}-
(1-\delta_{ij})\gamma_{i}\gamma_{j}\mathcal{F}_{i}\mathcal{F}_{j}\boldsymbol{G}(\boldsymbol{r}_{i},\boldsymbol{r}_{j})\label{eq:3-1}
\end{equation}
where $\gamma_{i}=6\pi\eta a_{i}$ are the self-frictions, $\boldsymbol{G}$
is a Green's function of the Stokes equation, $\boldsymbol{r}_{i}$
are the positions of the centers of the colloids and $\mathcal{F}_{i}=1+\frac{a_{i}^{2}}{6}\boldsymbol{\nabla}_{i}^{2}$
are the Faxen corrections that account for the finite radius, $a_{i}$,
of the colloids. This expression of the friction tensor is not limited
to the translationally invariant Green's function $8\pi\eta\boldsymbol{G}(\boldsymbol{r})=(\boldsymbol{\nabla}^{2}\boldsymbol{I}-\boldsymbol{\nabla}\boldsymbol{\nabla})\boldsymbol{r}$,
but holds for any Green's function, and is both positive definite
and symmetric. In an unbound fluid, the mutual friction tensors decay
inversely with distance and the rapidity of decay increases at the proximity
of boundaries. For sampling frequencies $\omega_{s}<<\gamma_{i}/m_{i}$,
the inertial term can be adiabatically eliminated from the Langevin
equation \cite{gardiner1984adiabatic,gardiner1985handbook} to arrive at the 
inertialess Langevin equation representing the positions of the particles.
After linearizing the motion about the line joining the centers of the particles, and decomposing into components parallel and perpendicular to the
separation vector, the parallel correlation functions in the frequency
domain are
\begin{equation}
C_{ii}^{\parallel}(\omega)=\frac{2K_{B}T(\mu^{\parallel}k_{j}^{2}DetM_{\parallel}+\mu^{\parallel}\omega^{2})}
{(DetA_{\parallel}-\omega^{2})^{2}+\omega^{2}(TrA_{\parallel})^{2}}\label{eq:4}
\end{equation}
and 
\begin{equation}
C_{ij}^{\parallel}(\omega)=\frac{2K_{B}T\mu_{ij}^{\parallel}(\omega^{2}-DetA_{\parallel})}
{(DetA_{\parallel}-\omega^{2})^{2}+\omega^{2}(TrA_{\parallel})^{2}}\label{eq:5}
\end{equation}
where $i,\,j=1,\,2$ and $i\neq j$. $A_{\parallel ij}=\mu_{ij}^{\parallel}k_{j}$
is the response matrix and $M_{\parallel ij}=\mu_{\parallel ij}$.
$\mu_{\parallel ij}$ is the inverse of the friction matrix.
Eqns.~\ref{eq:4} and \ref{eq:5} can be inverse Fourier transformed
to get auto and cross-correlations in the time domain. The auto-correlations
are given by (see details in the Supplementary Information)
\begin{align}
C_{ii}^{\parallel}(\tau) & =\frac{2K_{B}T}{\chi(k_{1}+k_{2})^{3}}\times\nonumber \\
 & \Bigg[\frac{\left(k_{j}^{2}\left(1-\frac{\mu_{ij}^{\parallel2}}{\mu_{ii}^{\parallel2}}\right)-
\frac{\left(k_{i}+k_{j}\right)^{2}\left(1-\chi\right)^{2}}{4}\right)\exp\left(-\beta_{-}\tau\right)}{1-\chi}+\nonumber \\
 & \frac{\left(\frac{\left(k_{i}+k_{j}\right)^{2}\left(1+\chi\right)^{2}}{4}-k_{j}^{2}
\left(1-\frac{\mu_{ij}^{\parallel2}}{\mu_{ii}^{\parallel2}}\right)\right)\exp\left(-\beta_{+}\tau\right)}{1+\chi}\Bigg]\label{eq:6}
\end{align}
whereas the cross-correlation is
\begin{gather}
C_{ij}^{\parallel}(\tau)=\frac{2K_{B}T\mu_{ij}^{\parallel}}{\sqrt{\gamma^{2}-\omega_{0}^{2}}}
\left[\exp\left(-\beta_{+}\tau\right)-\exp\left(-\beta_{-}\tau\right)\right]\label{eq:7}
\end{gather}
where 
\[
\chi=\sqrt{\left[1-\frac{4k_{1}k_{2}\left(1-\frac{\mu_{12}^{\parallel2}}
{\mu_{11}^{\parallel2}}\right)}{\left(k_{1}+k_{2}\right)^{2}}\right]}
\]
\[
\beta_{-}=\frac{\mu_{11}^{\parallel}\left(k_{1}+k_{2}\right)\left(1-\chi\right)}{2}
\]
\[
\beta_{+}=\frac{\mu_{1}^{\parallel}\left(k_{1}+k_{2}\right)\left(1+\chi\right)}{2}
\]
and $\gamma=TrA_{\parallel}$, $\omega_{0}^{2}=DetA_{\parallel}$.
Here we assumed $\mu_{11}^{\parallel}=\mu_{22}^{\parallel}$ and $\mu_{12}^{\parallel}=\mu_{21}^{\parallel}$
since the particles are identical. $\mu_{ii}^{\parallel}=\frac{1}{6\pi\eta a_{i}}$ and $\mu_{ij}^{\parallel}=\frac{1}{8\pi\eta a_{i}}\left(2-\frac{4a_{i}^{2}}{3r_{0}^{2}}\right)$,
$r_{0}$ is the center to center separation between the particles.
Clearly, the auto-correlations are double exponential functions and
the cross-correlation is the difference of two exponential functions.
It is clear that the thermal fluctuation of each
particle is statistically non-Markovian according to Doob's theorem. Note that this is the most general expressions for the auto and cross-correlation functions in contrast to that obtained in Ref.~\cite{meiners1999}, where the trap stiffnesses had been considered to be equal. Our results reduce to that reported in Ref.~\cite{meiners1999} for $k_1=k_2$.

We now explore the viscous and viscoelastic nature of the system, which will be manifested in the mean-square displacement (MSD) of the probe particle. This can be quantified using the equation \cite{tassieri2010measuring}
\begin{equation}
G^{*}(\omega)=\frac{k_{i}}{6\pi a_{i}}\left[\frac{2\left\langle R_{i}^{2}\right\rangle }{i\omega\left\langle \Delta\hat{R_{i}^{2}}(\omega)\right\rangle }-1\right]\label{eq:8}
\end{equation}
$G^{*}(\omega)$ is the frequency dependent dynamic complex modulus - 
the real part (elastic modulus) of which represents the amount of
energy store, while the imaginary part (viscous modulus) represents
dissipation of energy. $k_{i}$ is the stiffness of the i-th trap
in which the probe is confined, $a_{i}$ is the radius of the probe.
$\left\langle \Delta\hat{R_{i}^{2}}(\omega)\right\rangle $ is the
Fourier transform of the time dependent MSD of the thermal position
fluctuation of the probe particle. $\left\langle R_{i}^{2}\right\rangle $
is the time independent variance. For a single trapped particle in a viscous medium, the auto-correlation is given by
\begin{eqnarray}
G^{*}(\omega) & = & i\eta\omega\label{eq:9}
\end{eqnarray}
Hence, the surrounding medium where the single trapped particle is embedded
is purely dissipative in nature. Now, if we go through a similar process for a pair of particles assuming one particle to serve as a probe, then
\begin{alignat}{1}
G^{*}(\omega)= & \frac{k_{i}}{6\pi a_{i}}\times\Bigg[\omega^{2}\Biggl\{\frac{(A\beta_{-}+B\beta_{+})(A\beta_{+}+B\beta_{-})}{\omega^{2}(A\beta_{+}+B\beta_{-})^{2}
+\beta_{+}^{2}\beta_{-}^{2}(A+B)^{2}}\nonumber \\
 & -\frac{\beta_{+}\beta_{-}(A+B)^{2}}{\omega^{2}(A\beta_{+}+B\beta_{-})^{2}+\beta_{+}^{2}\beta_{-}^{2}(A+B)^{2}}\Biggr\}+\nonumber \\
 & i\omega\Biggl\{\frac{(A+B)(A\beta_{-}+B\beta_{+})\beta_{+}\beta_{-}}{\omega^{2}(A\beta_{+}+B\beta_{-})^{2}+
\beta_{+}^{2}\beta_{-}^{2}(A+B)^{2}}\label{eq:10}\\
 & +\frac{\omega^{2}(A+B)(A\beta_{+}+B\beta_{-})}{\omega^{2}(A\beta_{+}+B\beta_{-})^{2}+\beta_{+}^{2}\beta_{-}^{2}(A+B)^{2}}
\Biggr\}\Biggr]\nonumber 
\end{alignat}
where, A and B are the amplitudes of the exponential components of the auto-correlation function of the probe. From the above Eq.~\ref{eq:10}, it is very clear that the system whose properties are measured using the probe particle is not only dissipative but also stores energy as the elastic modulus is not zero but a function of frequency. We proceed to explore the variation of $G^{*}(\omega)$ as a function of trap stiffness ratio and particle separation at a fixed frequency of 400 Hz (close to our experimentally obtained trap corner frequency), the results of which are provided in Figs.~\ref{G_var}(a) and (b), respectively. It is interesting to observe that both $G'(\omega)$ and $G''(\omega)$ tend to increase with increasing stiffness ratio (Fig.~\ref{G_var}(a)), which implies that the trap stiffness of the probe particle needs be considerably greater than that of the other particle to obtain large $G^{*}(\omega)$. This can be attributed to the fact that a weak trap corresponds to large amplitude of Brownian fluctuations, which would in turn reduce the particle separation and thus lead to stronger hydrodynamic interaction (HI). For the same reason, as the particle separation is increased for a fixed trap stiffness ratio in (Fig.~\ref{G_var}(b)), the decrease in HI reduces $G^{*}(\omega)$, so that it approaches the single particle limit with $G'(\omega)=0$ and $G''(\omega)$ being a constant. 
\noindent\begin{minipage}[c]{\linewidth}%
 \centering %
\begin{minipage}[c]{0.4\linewidth}%
\begin{figure}[H]
\centering
\includegraphics[width=1\linewidth]{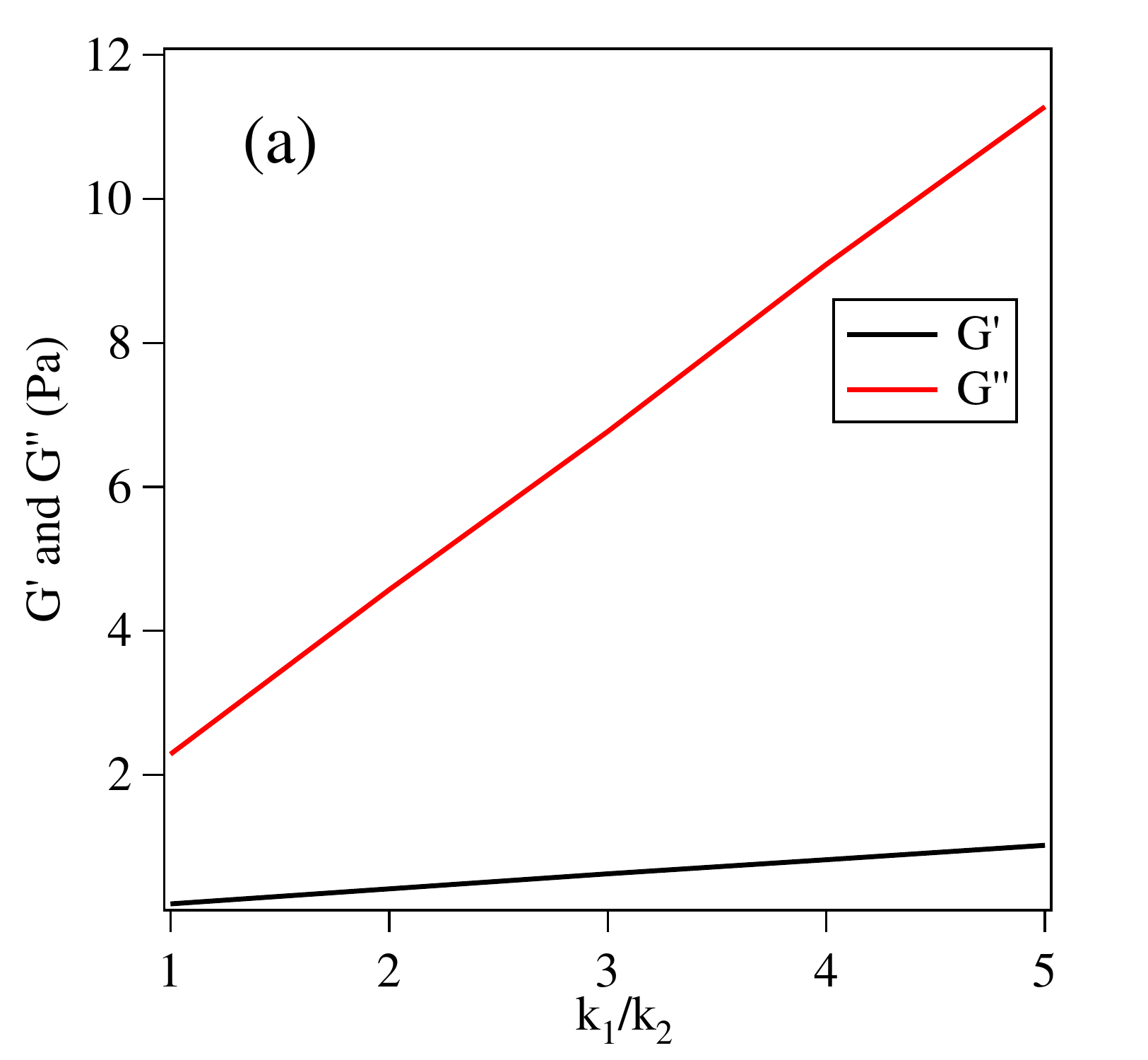} 
\end{figure}
\end{minipage}\hspace{0.05\linewidth} %
\begin{minipage}[c]{0.4\linewidth}%
\begin{figure}[H]
\centering
\includegraphics[width=1\linewidth]{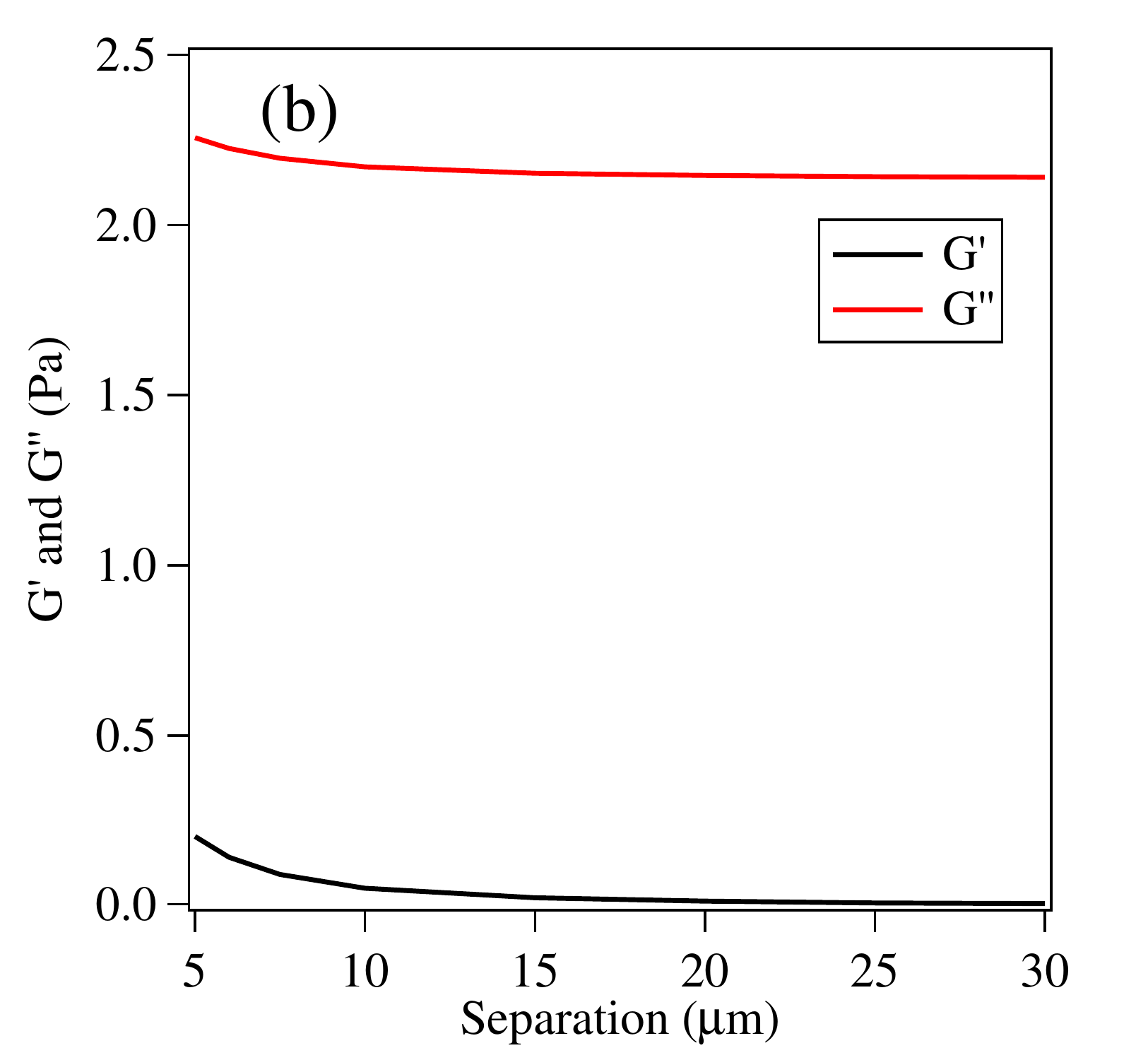} 
\end{figure}
\end{minipage}
\begin{minipage}[c]{1\linewidth}%
\begin{figure}[H]
\caption{\footnotesize{Plots of the real and imaginary part of $G^{*}$, $G'$ and $G''$ respectively w.r.t (a) trap stiffness ratio at fixed separation of 5 $\mu$m, (b) trap separation for fixed trap stiffnesses of 43 $\mu$N/m. Both plots are for frequency 400 Hz}}
\label{G_var} 
\end{figure}
\end{minipage}%
\end{minipage}

\emph{Experiment.} The experimental setup has been discussed in detail in Refs.~\cite{paul2017direct,paul2018two} - here we provide a brief sketch for completeness. We took care to avoid surface charges by using a 1 M NaCl-water solution as the viscous medium where the 3 $\mu$m
diameter polystyrene particles where embedded at very low volume fraction ($\phi\approx0.01$). The two optical traps were well-calibrated, and we kept them at initial separation of $5\pm0.1$ micron, and at a distance of 30 $\mu$m from the nearest wall. This distance is large enough to avoid optical cross-talk and the effects from surface charges  \cite{stilgoe2011phase}. The separation and individual laser powers were then varied to perform our experiment.  In order to ensure that the trapping beams did not influence each other, we measured the Brownian motion of one trapped particle in the presence of the other trap when the latter was still empty, and ensured that there were no changes in the properties of the Brownian fluctuations at the level of around 1\%. The normalized position fluctuation data for both particles were collected simultaneously for 10 s at a sampling frequency of 10 kHz. 

\emph{Results and discussions.} To check for consistency of our estimation of trap stiffness, we used a single trapped particle, and compared the stiffness obtained from the exponential fit to the time domain auto-correlation and that in the frequency domain where we fit the data to a lorentzian. The agreement was within 2\%, which thus provided confidence about our measurement process. For the two particle case, we first kept both the traps at the same stiffness and then varied the stiffness of one of the traps from 43 $\mu$N/m to 30 $\mu$N/m. The corresponding data is shown Fig.~\ref{stiff_v} (a)-(c). The data are in good agreement with the calculated ACF and CCF from Eq.~\ref{eq:6} and \ref{eq:7}, which are also plotted together to show the efficacy of the fit. Thus, it is clear that Doob's theorem is violated, and the resultant memory in the system is clearly a function of the stiffnesses of the two optical traps. After this, we proceeded to keep the trap stiffnesses equal, and varied the inter-particle separation. The corresponding ACF and CCF are shown in Fig.~\ref{dist_v}. The CCF dip reduces as the inter-particle distance increases since the hydrodynamic interaction also decreases proportionally. Concomitantly, the SNR deceases as well, so that the measurement deviates from the theoretical predictions (Fig.~\ref{dist_v}(c)).
\noindent\begin{minipage}[c]{1\linewidth}%
 \centering %
\begin{minipage}[c]{0.4\linewidth}%
 \begin{figure}[H]
\includegraphics[width=1\linewidth]{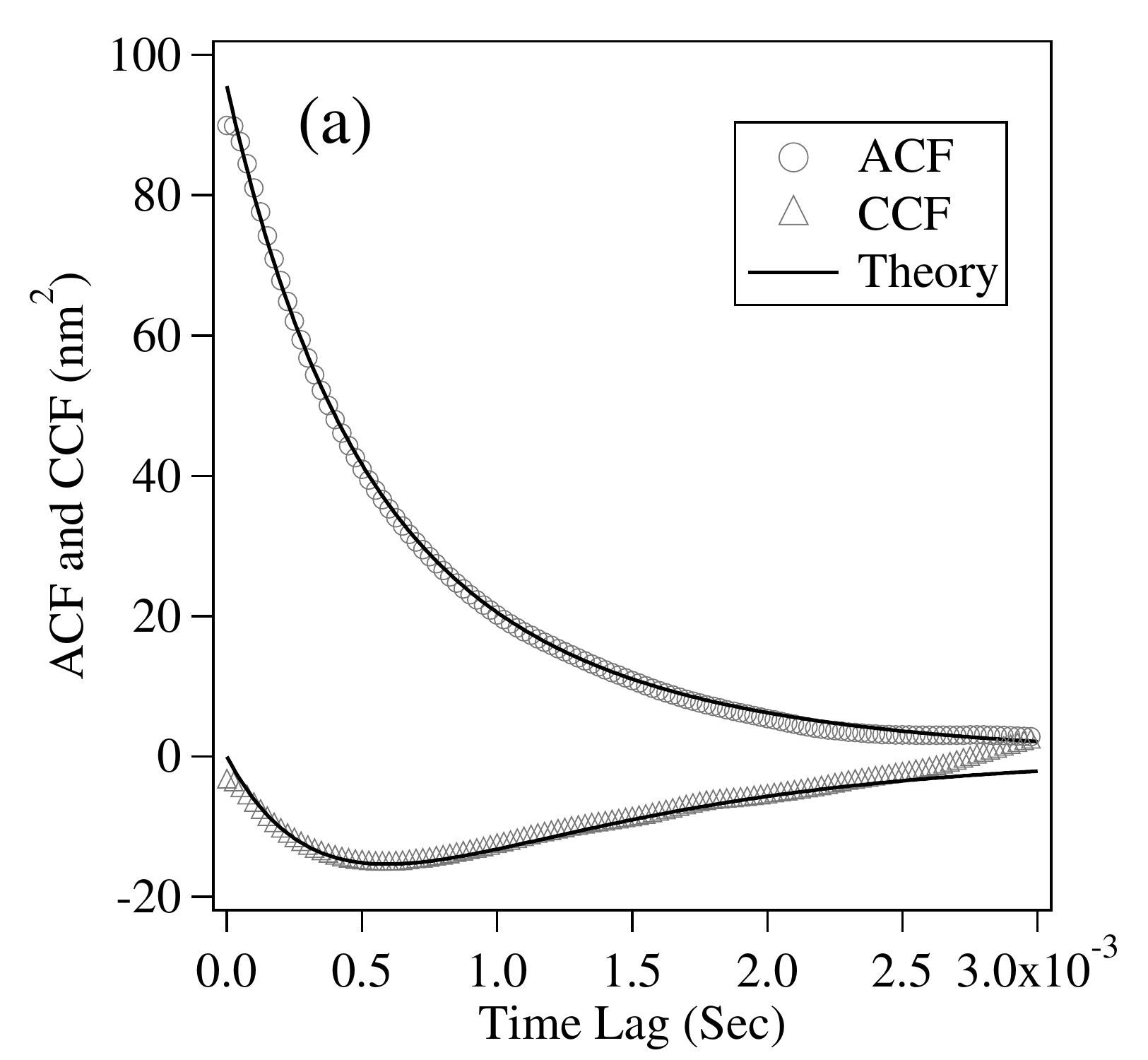} 
 \end{figure}
\end{minipage}
\begin{minipage}[c]{0.4\linewidth}%
 \begin{figure}[H]
\includegraphics[width=1\linewidth]{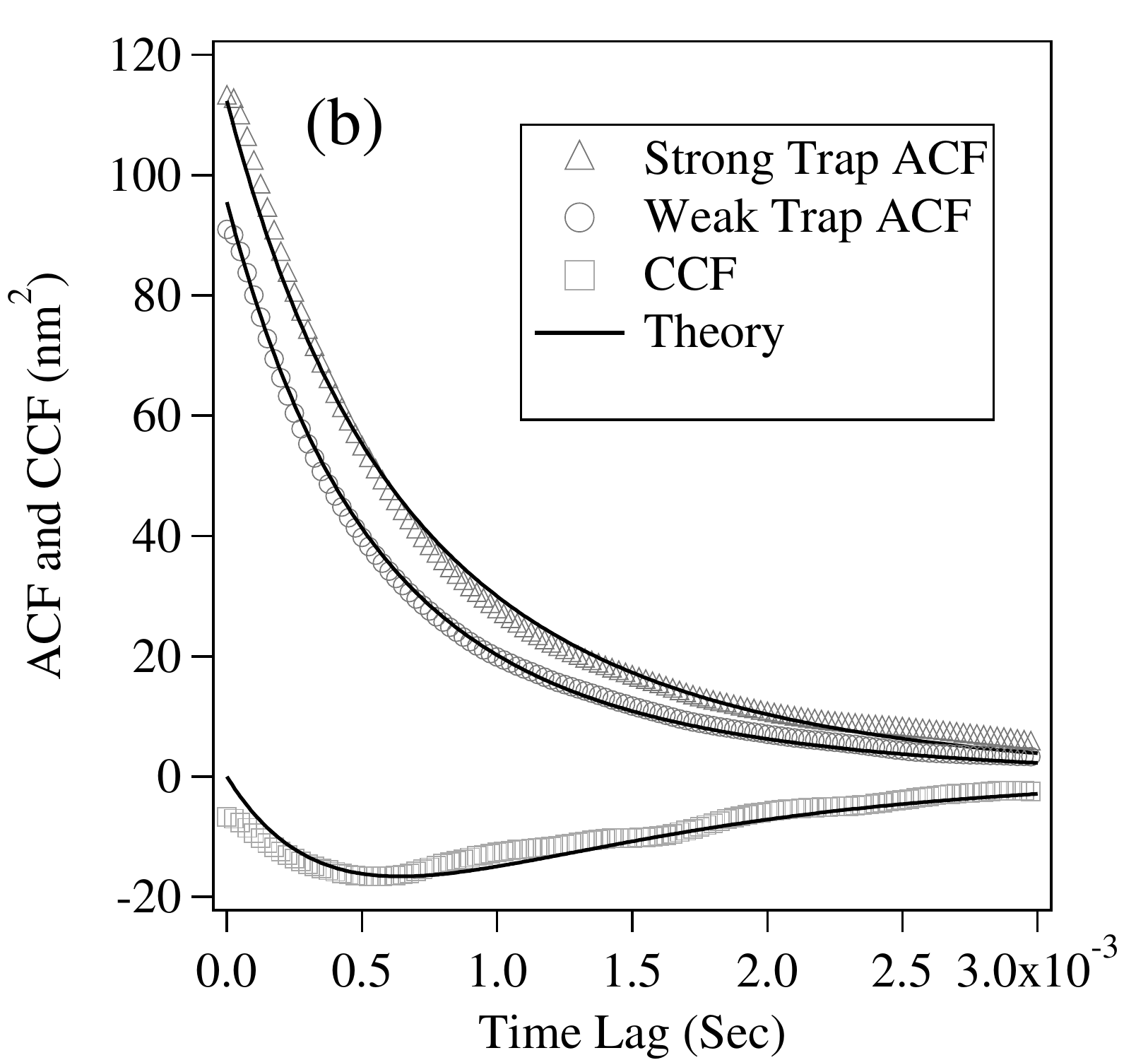} 
 \end{figure}
\end{minipage}\vspace{0.001\columnwidth}
\begin{minipage}[c]{0.4\linewidth}%
 \begin{figure}[H]
\includegraphics[width=1\linewidth]{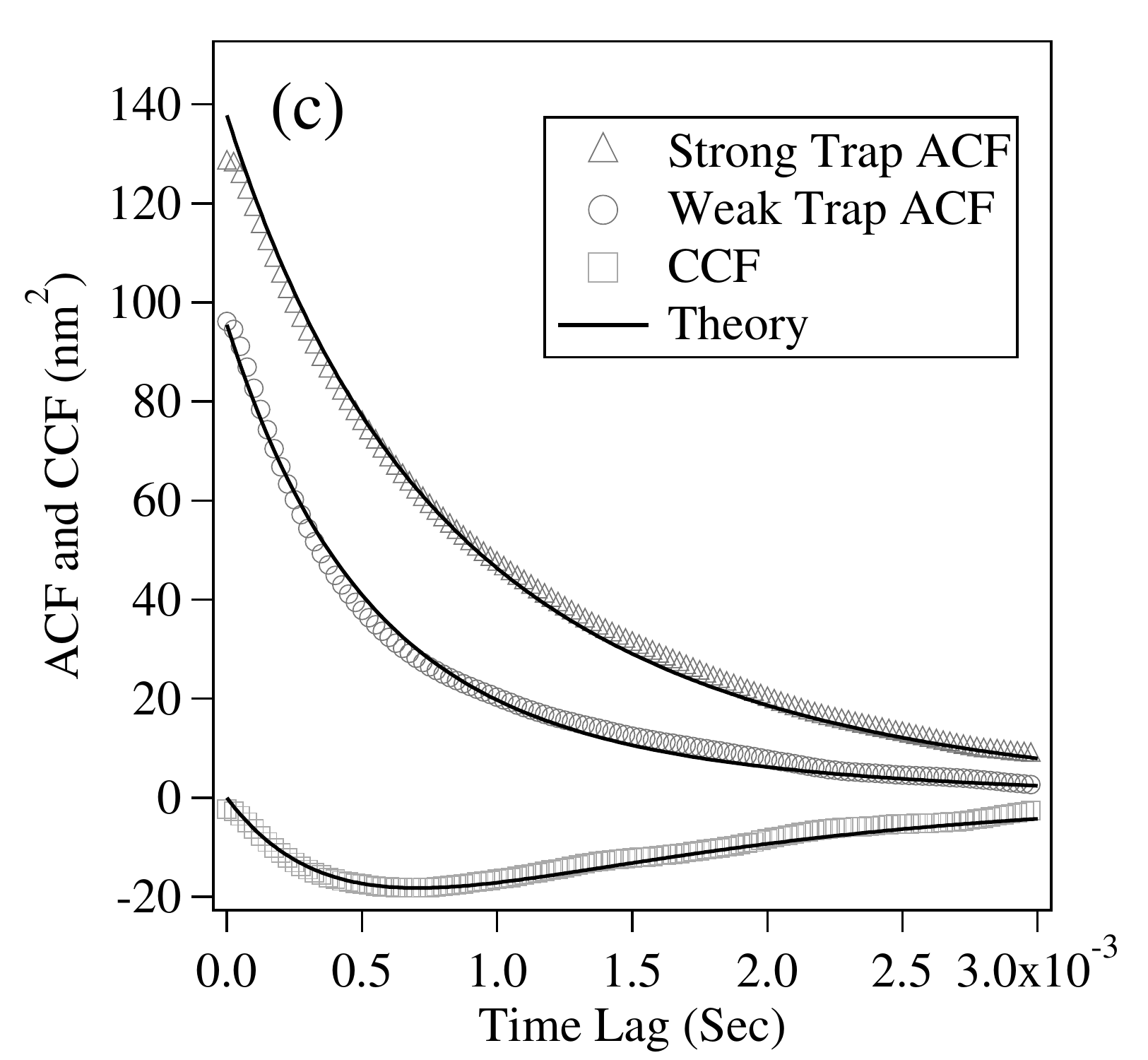}  
\end{figure}
\end{minipage}
 \hspace{0.01\linewidth}
\begin{minipage}[c]{0.4\linewidth}%
\begin{figure}[H]
\caption{\footnotesize{Auto-correlation function (ACF) and cross-correlation function (CCF)
w.r.t time lag at particle separation 5 $\mu$m., with $k_{1}=43$ and (a) $k_{2}=43$ $\mu$N/m, (b) $k_{2}=36.5$ $\mu$N/m, and (c) $k_{2}=30$ $\mu$N/m}}
\label{stiff_v}
\end{figure}
\end{minipage}%
\end{minipage} 
\noindent\begin{minipage}[c]{1\linewidth}%
 \centering %
\begin{minipage}[c]{0.4\linewidth}%
\begin{figure}[H]
\includegraphics[width=1\linewidth]{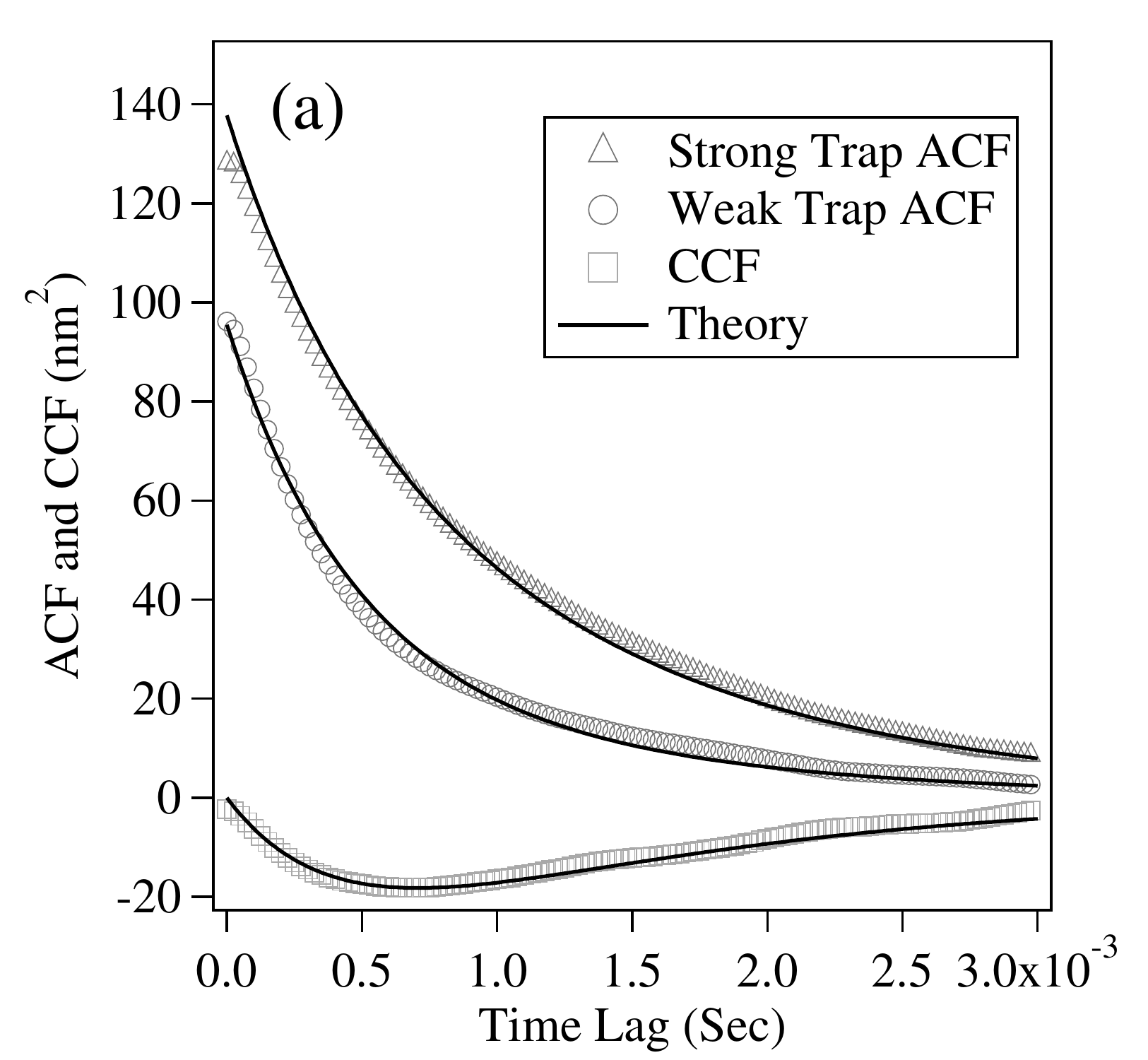} 
\end{figure}
\end{minipage}\hspace{0.05\linewidth} %
\begin{minipage}[c]{0.4\linewidth}%
\begin{figure}[H]
\includegraphics[width=1\linewidth]{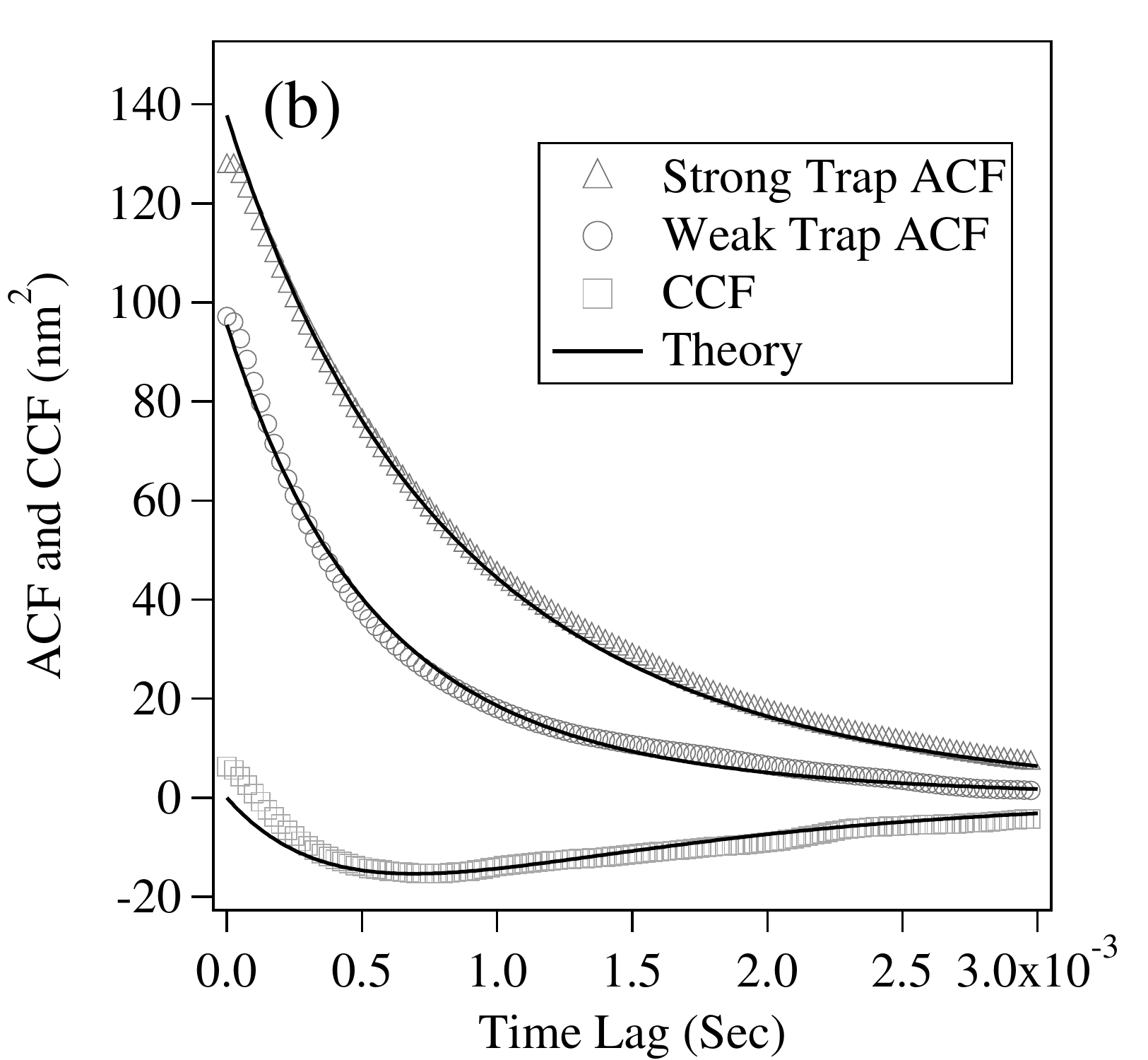} 
\end{figure}
\end{minipage}\vspace{0.001\linewidth}
\begin{minipage}[c]{0.4\linewidth}%
\begin{figure}[H]
\includegraphics[width=1\linewidth]{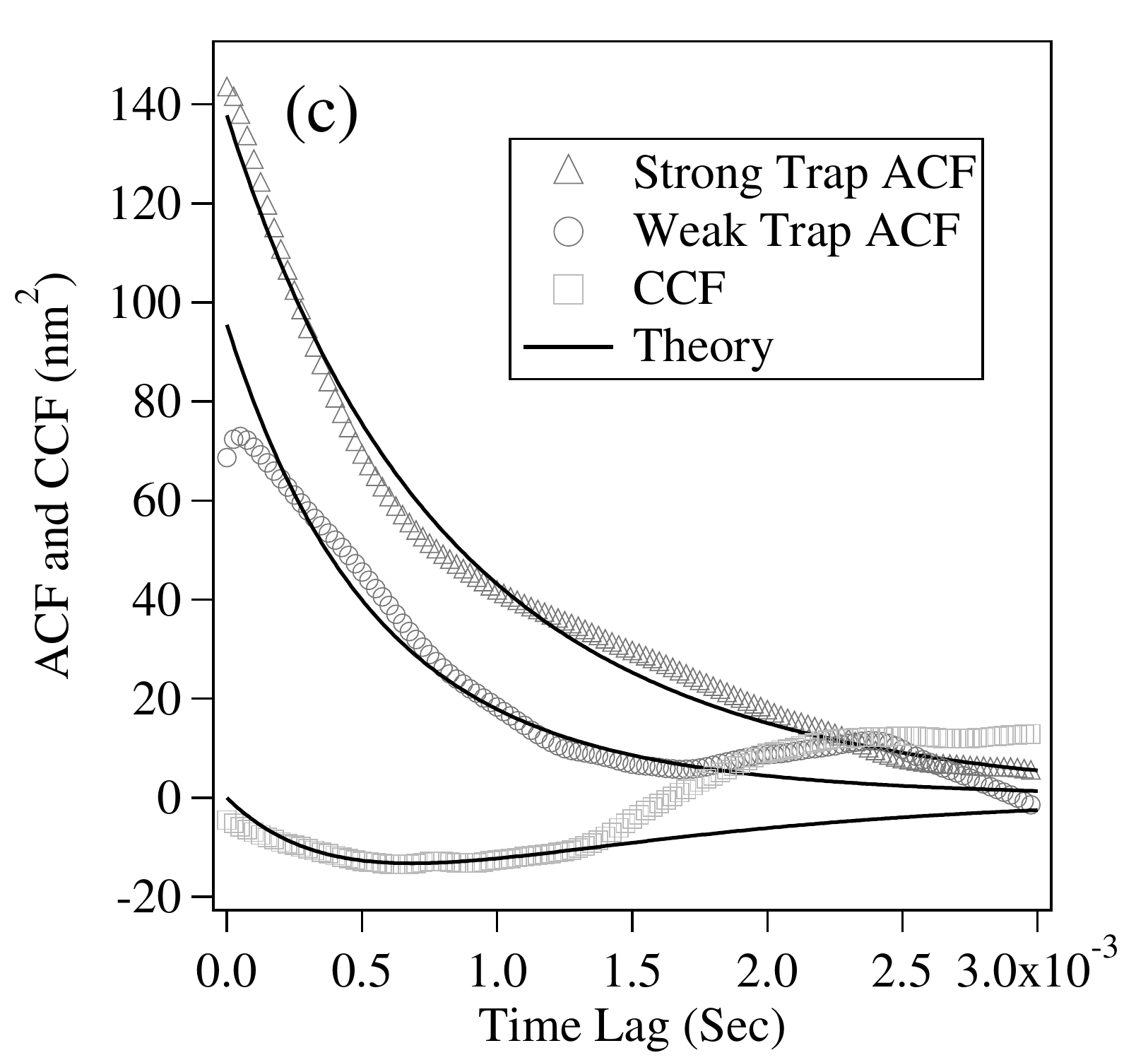} 
\end{figure}
\end{minipage}
\hspace{0.06\linewidth}
\begin{minipage}[c]{0.4\linewidth}%
\begin{figure}[H]
\caption{\footnotesize{ACF and CCF w.r.t time lag. $k_{1}=43$
$\mu$N/m and $k_{2}=30$, with inter-particle separation (a) 5 $\mu$m (b) 
6 $\mu$m, and (c) 7 $\mu$m.}}
\label{dist_v} 
\end{figure}
\end{minipage}%
\end{minipage}

To determine the energy stored in the system, we use one of the traps
(with k=43 $\mu$N/m) as probe, and use Eqn. \ref{eq:8} to calculate the
real and imaginary part of  $G^{*}$ from the data. We fourier
transform the calculated MSD using the method reported in Ref.~
\cite{evans2009direct}, and substitute it in Eqn. \ref{eq:8}. We
use standard data smoothing techniques to generate $G'$ and $G''$
which are shown in Fig.~\ref{G}. It is clear that the experimental
measurement matches the theoretical prediction (Eqn. \ref{eq:10})
rather well. Fig.~\ref{G}(a) shows the data for an isolated single
trapped particle, from which it is apparent that $G'$ is independent
of frequency. As per theory, the value of $G'$ should be zero, but
the slight offset observed in the experimental data is due to noise artifacts.
On the other hand, $G''$ increases linearly as expected for a viscous medium. Fig.~\ref{G}(b) and (c) show data for the coupled system with k1=k2=43 $\mu$N/m,
and k1=43 $\mu$N/m and k2=36.5 $\mu$N/m, respectively. The value of $G'$ is now three orders of magnitude higher than that for the single particle case, which quantitatively demonstrates how the two-particle system acquires viscoelastic properties, which are manifested in both the magnitude and frequency dependence of $G'$.
\noindent\begin{minipage}[c]{\linewidth}%
 \centering %
\begin{minipage}[c]{0.4\linewidth}%
\begin{figure}[H]
\includegraphics[width=1\linewidth]{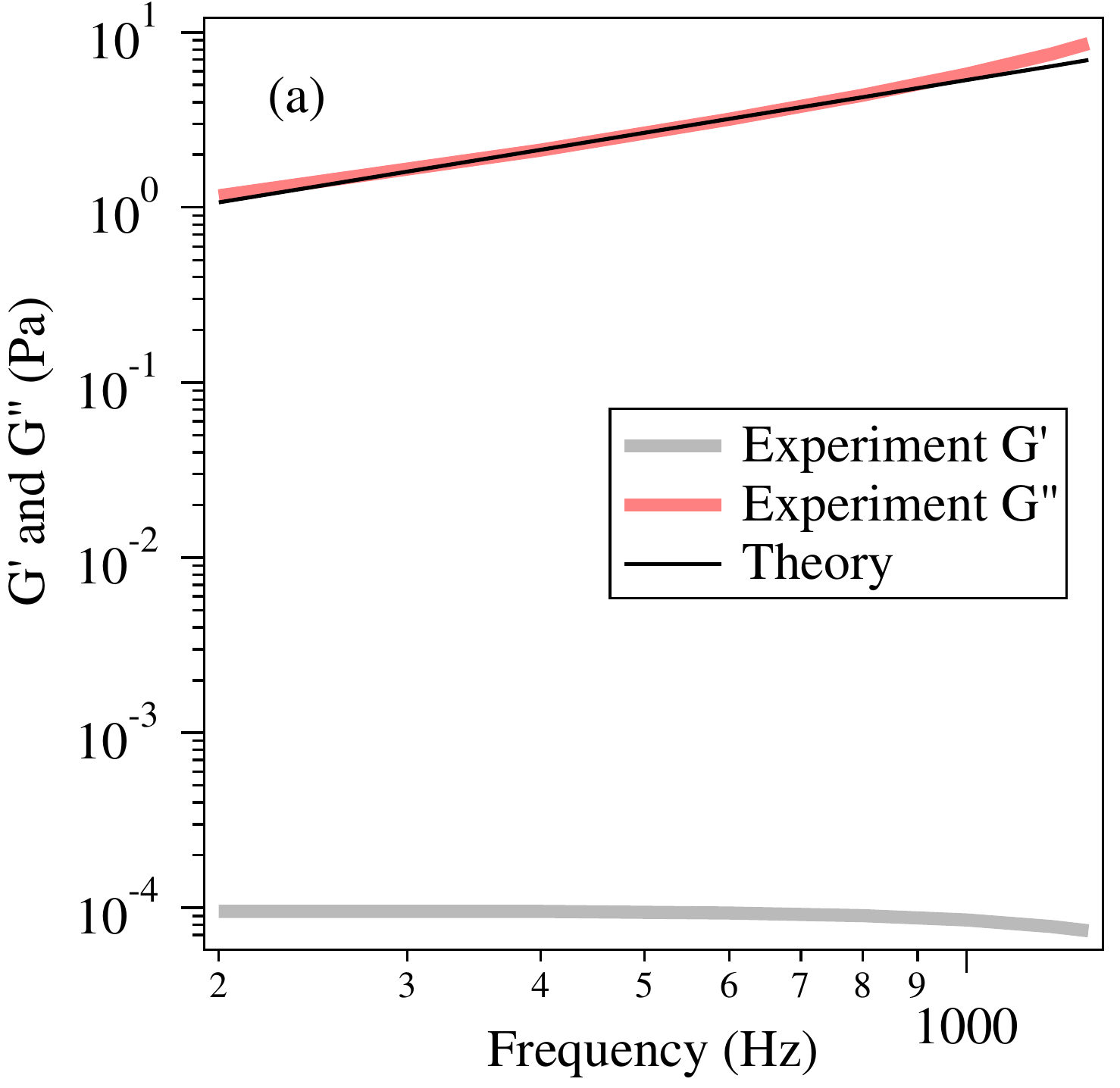} 
\end{figure}
\end{minipage}\hspace{0.05\linewidth} %
\begin{minipage}[c]{0.4\linewidth}%
\begin{figure}[H]
\includegraphics[width=1\linewidth]{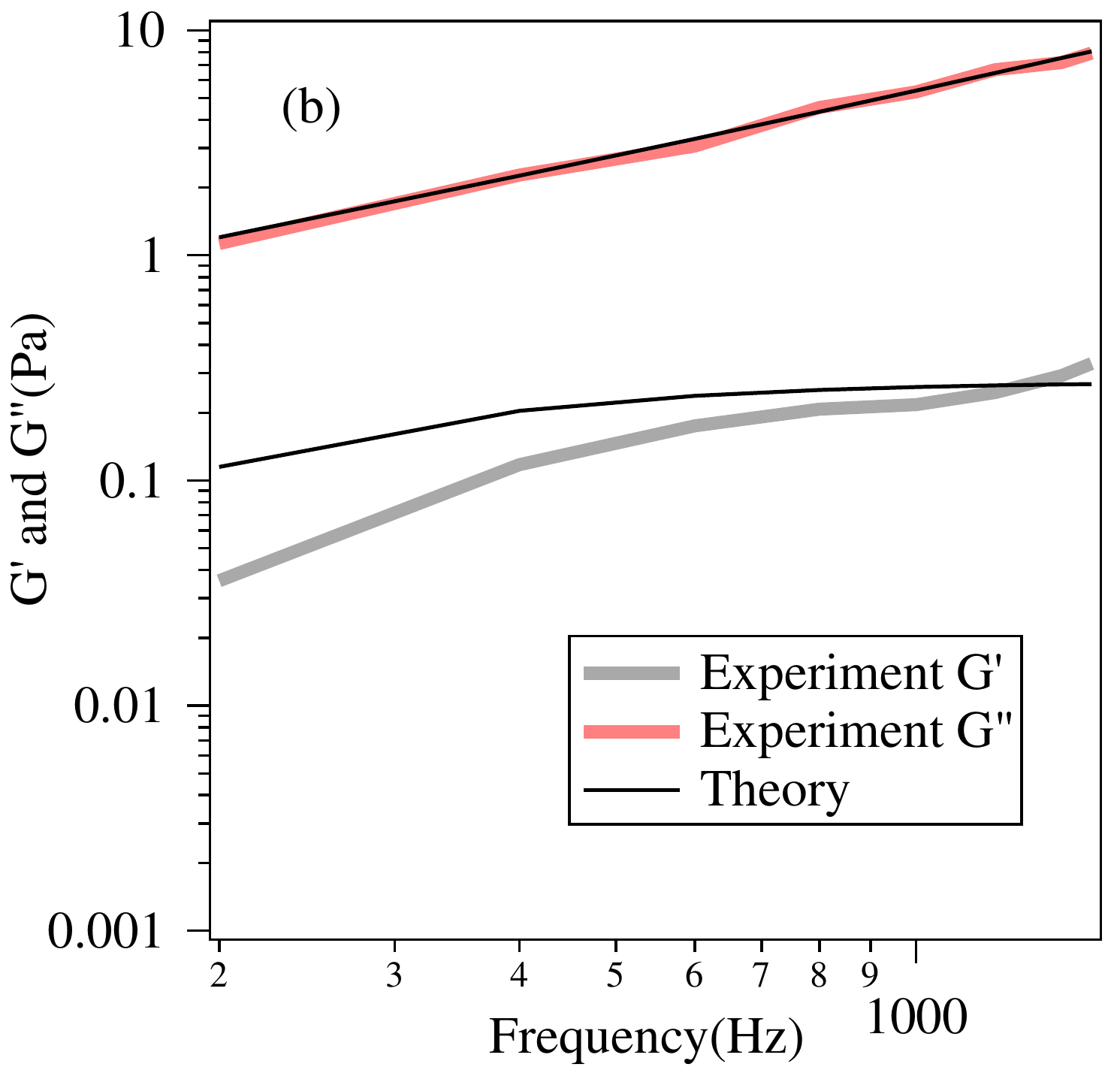} 
\end{figure}
\end{minipage}\vspace{0.002\linewidth}
\begin{minipage}[c]{0.4\linewidth}%
\begin{figure}[H]
\includegraphics[width=1\linewidth]{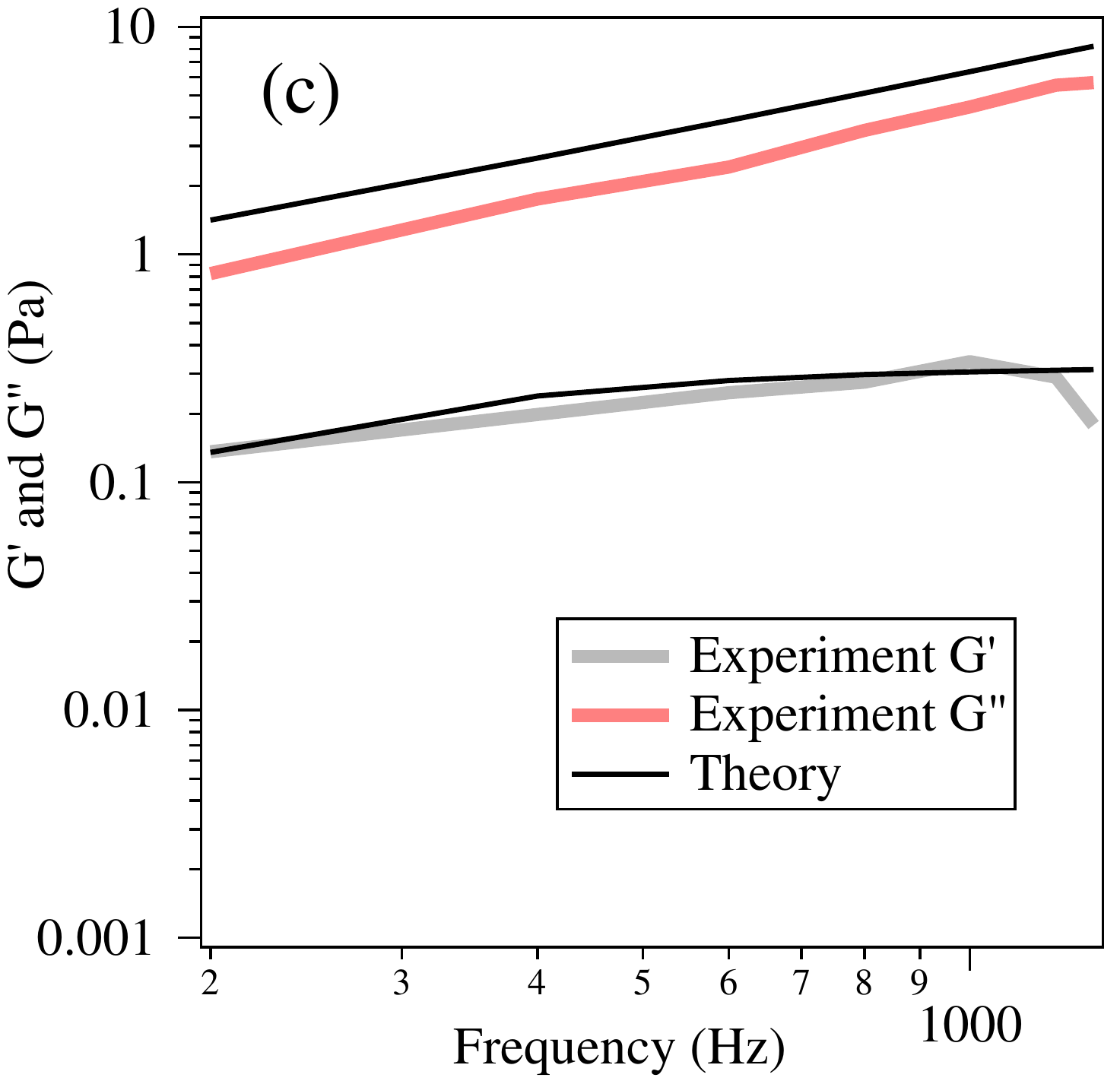} 
\end{figure}
\end{minipage}\hspace{0.05\linewidth} %
\begin{minipage}[c]{0.4\linewidth}%
\begin{figure}[H]
\caption{\footnotesize{Plots of the real and imaginary part of $G^{*}$, $G'$ and $G''$ respectively w.r.t frequency. (a) single trapped particle ($k=43$ $\mu$N/m), (b) for
$k_{1}=k_{2}=43$ $\mu$N/m, (c) for $k_{1}=43$ $\mu$N/m, $k_{2}=37.7$
$\mu$N/m.}}
\label{G} 
\end{figure}
\end{minipage}%
\end{minipage}

In conclusion, we demonstrate comprehensively that non-Markovian behaviour
of a system can be interpreted as the simple superposition of underlying
uncorrelated Markovian processes that are coupled by an interaction.
We choose two Brownian particles in separate optical traps that are
viscously coupled, so that the resultant system mimics a
single trapped particle in a viscoelastic media. A superposition of
the individual time constants of the optical traps appear in the auto-correlation function of each particle, so that they are double exponential in nature and similar to a single particle optically trapped in viscoelastic media. Thus, an apparently memory-less process develops
memory, which gradually disappears as the hydrodynamic coupling between
the particles reduces when their separation is increased or when the optical trap stiffnesses are skewed. The emergence of memory also implies storage of energy in the system, and we directly quantify this by measuring the equivalent storage and loss moduli - that makes the analogy with viscoelastic media even clearer. Note that the system here obeys detailed balance\cite{paul2017} - an obvious extension would be to investigate the behaviour and evolution of the
energy stored in this system when it is driven out of equilibrium
by the use of a temperature gradient or in the presence of convective
flows that can be generated experimentally \cite{roy2016}. We are
presently pursuing these directions.

This work was supported by the Indian Institute of Science Education
and Research, Kolkata, funded by the Ministry of Human Resource Development,
Govt. of India. We acknowledge Dr. R. Adhikari of The Institute of
Mathematical Sciences for help in developing the theoretical formalism.

\bibliographystyle{apsrev4-1}
%

\end{document}


\title{Supplementary Information}

\author{Shuvojit Paul}

\affiliation{Indian Institute of Science Education and Research, Kolkata }

\author{Randhir Kumar}

\affiliation{Indian Institute of Science Education and Research, Kolkata }

\author{Ayan Banerjee}
\email{ayan@iiserkol.ac.in}

\affiliation{Indian Institute of Science Education and Research, Kolkata }
\maketitle

\section{Theory}

Here we outline the major steps in deriving the correlation functions
and the complex shear modulus on the Smoluchowski time scale, i.e.
the over-damped regime, starting from Langevin equations

\begin{equation}
m_{i}\dot{\boldsymbol{v}}_{i}=-\boldsymbol{\gamma}_{ij}\boldsymbol{v}_{j}-{\bf \boldsymbol{\nabla}}_{i}U+{\bf \boldsymbol{\xi}}_{i}\label{eq:1}
\end{equation}
 
\begin{equation}
{\bf \dot{R}}_{i}=\boldsymbol{v}_{i}\label{eq:2}
\end{equation}

presented and explained in the main text. Due to the experimental
limitation on the data sampling frequency, we can assume the
momentum to be rapidly relaxing on the time scale of the trap motion.
Therefore, we can adiabatically eliminate the momentum \cite{gardiner1984adiabatic,gardiner1985handbook}
and get,

\begin{equation}
\begin{array}{c}
\boldsymbol{\gamma}_{ij}({\bf R}_{i}^{0},{\bf R}_{j}^{0}){\bf \dot{R}}_{j}+{\bf \boldsymbol{\nabla}}_{i}U={\bf \boldsymbol{\xi}}_{i}\end{array}\label{eq:3}
\end{equation}
\begin{gather*}
\langle\boldsymbol{\xi}_{i}(t)\rangle=\boldsymbol{0}\\
\left\langle \boldsymbol{\xi}_{i}(t)\boldsymbol{\xi}_{j}(t')\right\rangle =2k_{B}T\boldsymbol{\gamma}_{ij}({\bf R}_{i}^{0},{\bf R}_{j}^{0})\delta(t-t')
\end{gather*}
Since the trap-center separation is far greater than the standard
deviation of the position fluctuations of individual particles, $\boldsymbol{\gamma}_{ij}$
can be considered time-independent and a function of the trap-center
separation. Equation \ref{eq:3} can be inverted and presented in
terms of approximate mobility tensors $\boldsymbol{\mu}_{ij}({\bf R}_{i}^{0},{\bf R}_{j}^{0})=\boldsymbol{\gamma}_{ij}^{-1}({\bf R}_{i}^{0},{\bf R}_{j}^{0})$
in the following manner:

\begin{equation}
\begin{array}{c}
{\bf \dot{R}}_{i}+\boldsymbol{\mu}_{ij}{\bf \boldsymbol{\nabla}}_{j}U=\boldsymbol{\mu}_{ij}{\bf \boldsymbol{\xi}}_{j}\end{array}\label{eq:4}
\end{equation}
\[
\frac{d}{dt}\begin{bmatrix}{\bf R}_{1}\\
{\bf R}_{2}
\end{bmatrix}=-\begin{bmatrix}\mu k_{1}\boldsymbol{{\bf \delta}} & \boldsymbol{{\bf \mu}}_{12}k_{2}\\
\boldsymbol{{\bf \mu}}_{21}k_{1} & \mu k_{2}{\bf \boldsymbol{\delta}}
\end{bmatrix}\begin{bmatrix}{\bf R}_{1}\\
{\bf R}_{2}
\end{bmatrix}+\begin{bmatrix}\mu{\bf \boldsymbol{\delta}} & \boldsymbol{{\bf \mu}}_{12}\\
\boldsymbol{{\bf \mu}}_{21} & \mu\boldsymbol{\delta}
\end{bmatrix}\begin{bmatrix}{\bf \boldsymbol{\xi}}_{1}\\
{\bf \boldsymbol{\xi}}_{2}
\end{bmatrix}
\]
where $\boldsymbol{\mu}_{11}=\mu\boldsymbol{\delta}=\boldsymbol{\mu}_{22}$
as the two particles are identical. The steady-state solution of equation
\ref{eq:4} in frequency space is derived easily by a Fourier transformation
as
\begin{align}
{\bf R}(\omega)=(-i\omega\boldsymbol{{\bf \delta}}+{\bf A})^{-1}{\bf M\boldsymbol{\xi}}(\omega)\label{eq:5}
\end{align}
where ${\bf A}=\begin{bmatrix}\mu k_{1}{\bf \boldsymbol{\delta}} & \boldsymbol{{\bf \mu}}_{12}k_{2}\\
\boldsymbol{{\bf \mu}}_{21}k_{1} & \mu k_{2}\boldsymbol{{\bf \delta}}
\end{bmatrix}$ and ${\bf M}=\begin{bmatrix}\mu\boldsymbol{{\bf \delta}} & \boldsymbol{{\bf \mu}}_{12}\\
\boldsymbol{{\bf \mu}}_{21} & \mu\boldsymbol{\delta}
\end{bmatrix}$. ${\bf \boldsymbol{\delta}}$ is $3\times3$ unit matrix.

\subsubsection{Correlation functions}

The correlation matrix can be written as
\begin{eqnarray*}
\langle{\bf R}(\omega){\bf R}^{\dagger}(\omega)\rangle= & (-i\omega{\bf \boldsymbol{\delta}}+{\bf A})^{-1}{\bf M}\langle\boldsymbol{{\bf \xi}}(\omega){\bf \boldsymbol{\xi}}^{\dagger}(\omega)\rangle{\bf M}(i\omega{\bf \boldsymbol{\delta}}+{\bf A}^{T})^{-1}
\end{eqnarray*}

\begin{eqnarray}
\frac{1}{2k_{B}T}\boldsymbol{C}_{\Delta\Delta}=(-i\omega\boldsymbol{{\bf \delta}}+{\bf A})^{-1}{\bf M}(i\omega\boldsymbol{{\bf \delta}}+{\bf A}^{T})^{-1}\label{eq:6}
\end{eqnarray}

Since, ${\bf M}\langle\boldsymbol{{\bf \xi}}(\omega){\bf \boldsymbol{\xi}}^{\dagger}(\omega)\rangle{\bf M}=2k_{B}T\times{\bf M}$.
Now, for a given experimental set-up, the particles' motion can be
decomposed into components parallel and perpendicular to the trap-separation.
This $\boldsymbol{C}_{\Delta\Delta}$ can be decomposed into $\boldsymbol{C}_{\Delta\Delta}^{\parallel}$
and $\boldsymbol{C}_{\Delta\Delta}^{\perp}$, which correspond to
motion along the trap separation and perpendicular to it, respectively.
Here, we are interested only in the parallel component because the
experimental detection system is oriented in that direction. Considering
remote boundaries, the translational symmetry of the friction tensor
can be used to express it in the following manner: 
\[
\boldsymbol{\gamma}_{ij}=\gamma_{ij}^{\parallel}(\boldsymbol{r}_{0})\hat{\boldsymbol{r}}_{0}\hat{\boldsymbol{r}}_{0}+\gamma_{ij}^{\perp}(\boldsymbol{r}_{0})(\boldsymbol{I}-\hat{\boldsymbol{r}}_{0}\hat{\boldsymbol{r}}_{0})
\]

Here, $\boldsymbol{r}_{0}$ is the trap-center separation, $\gamma_{ij}^{\parallel}(\boldsymbol{r}_{0})$
is the friction coefficient for the motion parallel to $\boldsymbol{r}_{0}$
and $\gamma_{ij}^{\perp}(\boldsymbol{r}_{0})$ is the same but for
the motion perpendicular to the trap-center separation. Corresponding
mobility matrices are given by $\gamma_{ik}^{\parallel}\mu_{kj}^{\parallel}=\delta_{ij}$.
Now,\begin{widetext}

\begin{multline*}
\boldsymbol{C}_{\Delta\Delta}^{\parallel}=(-i\omega{\bf \boldsymbol{\delta}}+{\bf A})_{\parallel}^{-1}{\bf M}_{\parallel}(i\omega{\bf \boldsymbol{\delta}}+{\bf A}_{\parallel}^{T})^{-1}=\frac{1}{(DetA_{\parallel}-\omega^{2})^{2}+\omega^{2}(TrA_{\parallel})^{2}}\\
\times\begin{bmatrix}\mu^{\parallel}k_{2}-i\omega & -\mu_{12}^{\parallel}k_{2}\\
-\mu_{21}^{\parallel}k_{1} & \mu^{\parallel}k_{1}-i\omega
\end{bmatrix}\begin{bmatrix}\mu^{\parallel} & \mu_{12}^{\parallel}\\
\mu_{21}^{\parallel} & \mu^{\parallel}
\end{bmatrix}\begin{bmatrix}\mu^{\parallel}k_{2}+i\omega & -\mu_{12}^{\parallel}k_{2}\\
-\mu_{21}^{\parallel}k_{1} & \mu^{\parallel}k_{1}+i\omega
\end{bmatrix}\\
=\frac{1}{(DetA_{\parallel}-\omega^{2})^{2}+\omega^{2}(TrA_{\parallel})^{2}}\times\begin{bmatrix}\mu^{\parallel}k_{2}^{2}DetM_{\parallel}+\mu^{\parallel}\omega^{2} & -(DetA_{\parallel}-\omega^{2})\mu_{21}^{\parallel}\\
-(DetA_{\parallel}-\omega^{2})\mu_{12}^{\parallel} & \mu^{\parallel}k_{1}^{2}DetM_{\parallel}+\mu^{\parallel}\omega^{2}
\end{bmatrix}
\end{multline*}

So, 
\[
\begin{bmatrix}C_{11}^{\parallel} & C_{12}^{\parallel}\\
C_{21}^{\parallel} & C_{22}^{\parallel}
\end{bmatrix}=\frac{2K_{B}T}{(DetA_{\parallel}-\omega^{2})^{2}+\omega^{2}(TrA_{\parallel})^{2}}\begin{bmatrix}\mu^{\parallel}k_{2}^{2}DetM_{\parallel}+\mu^{\parallel}\omega^{2} & -(DetA_{\parallel}-\omega^{2})\mu_{21}^{\parallel}\\
-(DetA_{\parallel}-\omega^{2})\mu_{12}^{\parallel} & \mu^{\parallel}k_{1}^{2}DetM_{\parallel}+\mu^{\parallel}\omega^{2}
\end{bmatrix}
\]

\end{widetext}

The auto-correlations of the Brownian position fluctuations of the
particles in the traps of stiffnesses $k_{1}$ and $k_{2}$ in the
frequency domain are given by 
\begin{equation}
C_{11}(\omega)=\frac{2K_{B}T(\mu^{\parallel}k_{2}^{2}DetM_{\parallel}+\mu^{\parallel}\omega^{2})}{(DetA_{\parallel}-\omega^{2})^{2}+\omega^{2}(TrA_{\parallel})^{2}}\label{eq:8}
\end{equation}

\begin{equation}
C_{22}(\omega)=\frac{2K_{B}T(\mu^{\parallel}k_{1}^{2}DetM_{\parallel}+\mu^{\parallel}\omega^{2})}{(DetA_{\parallel}-\omega^{2})^{2}+\omega^{2}(TrA_{\parallel})^{2}}\label{eq:9}
\end{equation}

respectively and 
\begin{equation}
C_{12}(\omega)=C_{21}(\omega)=\frac{2K_{B}T\mu_{21}^{\parallel}(\omega^{2}-DetA_{\parallel})}{(DetA_{\parallel}-\omega^{2})^{2}+\omega^{2}(TrA_{\parallel})^{2}}\label{eq:10}
\end{equation}

is the representation of the cross-correlation function in frequency
domain. 

\begin{equation}
\begin{array}{c}
DetA_{\parallel}=k_{1}k_{2}(\mu_{11}^{\parallel}\mu_{22}^{\parallel}-\mu_{12}^{\parallel}\mu_{21}^{\parallel})\\
TrA_{\parallel}=k_{1}\mu_{11}^{\parallel}+k_{2}\mu_{22}^{\parallel}
\end{array}\label{eq:11}
\end{equation}

Equations \ref{eq:8}, \ref{eq:9} and \ref{eq:10} can be inverse
Fourier transformed to get auto and cross-correlations in the time
domain. The auto-correlations are given by

\begin{align}
C_{11}^{\parallel}(\tau) & =\frac{2K_{B}T}{\chi(k_{1}+k_{2})^{3}}\times\nonumber \\
 & \Bigg[\frac{\left(k_{2}^{2}\left(1-\frac{\mu_{12}^{\parallel2}}{\mu_{11}^{\parallel2}}\right)-\frac{\left(k_{1}+k_{2}\right)^{2}\left(1-\chi\right)^{2}}{4}\right)\exp\left(-\beta_{-}\tau\right)}{1-\chi}+\nonumber \\
 & \frac{\left(\frac{\left(k_{1}+k_{2}\right)^{2}\left(1+\chi\right)^{2}}{4}-k_{2}^{2}\left(1-\frac{\mu_{12}^{\parallel2}}{\mu_{11}^{\parallel2}}\right)\right)\exp\left(-\beta_{+}\tau\right)}{1+\chi}\Bigg]\label{eq:12_1}
\end{align}

\begin{align}
C_{22}^{\parallel}(\tau) & =\frac{2K_{B}T}{\chi(k_{1}+k_{2})^{3}}\times\nonumber \\
 & \Biggl[\frac{\left(k_{1}^{2}\left(1-\frac{\mu_{12}^{\parallel2}}{\mu_{11}^{\parallel2}}\right)-\frac{\left(k_{1}+k_{2}\right)^{2}\left(1-\chi\right)^{2}}{4}\right)\exp\left(-\beta_{-}\tau\right)}{1-\chi}+\nonumber \\
 & \frac{\left(\frac{\left(k_{1}+k_{2}\right)^{2}\left(1+\chi\right)^{2}}{4}-k_{1}^{2}\left(1-\frac{\mu_{12}^{\parallel2}}{\mu_{11}^{\parallel2}}\right)\right)\exp\left(-\beta_{+}\tau\right)}{1+\chi}\Biggr]\label{eq:13_1}
\end{align}

and the cross-correlation is

\begin{align}
C_{12}^{\parallel}(\tau) & =C_{21}^{\parallel}(\tau)=\frac{2K_{B}T\mu_{12}^{\parallel}}{\sqrt{\gamma^{2}-\omega_{0}^{2}}}\left[\exp\left(-\beta_{+}\tau\right)-\exp\left(-\beta_{-}\tau\right)\right]\label{eq:14_1}
\end{align}

where 
\[
\chi=\sqrt{\left[1-\frac{4k_{1}k_{2}\left(1-\frac{\mu_{12}^{\parallel2}}{\mu_{11}^{\parallel2}}\right)}{\left(k_{1}+k_{2}\right)^{2}}\right]}
\]

\[
\beta_{-}=\frac{\mu_{11}^{\parallel}\left(k_{1}+k_{2}\right)\left(1-\chi\right)}{2}
\]

\[
\beta_{+}=\frac{\mu_{1}^{\parallel}\left(k_{1}+k_{2}\right)\left(1+\chi\right)}{2}
\]

and $\gamma=TrA_{\parallel}$, $\omega_{0}^{2}=DetA_{\parallel}$.
Here we assumed $\mu_{11}^{\parallel}=\mu_{22}^{\parallel}$ and $\mu_{12}^{\parallel}=\mu_{21}^{\parallel}$
since the particles are identical. $\mu_{ii}^{\parallel}=\frac{1}{6\pi\eta a_{i}}$
and $\mu_{ij}^{\parallel}=\frac{1}{8\pi\eta a_{i}}\left(2-\frac{4a_{i}^{2}}{3r_{0}^{2}}\right)$. 

\subsubsection{Complex shear modulus}

Now, if we consider one of the two particles as probe to investigate the
viscous and elastic nature of the system, then this
will be manifested by the mean-square displacement (MSD) of it's Brownian
fluctuation through the equation given below \cite{tassieri2010measuring}.
\begin{equation}
G^{*}(\omega)=\frac{k_{i}}{6\pi a_{i}}\left[\frac{2\left\langle R_{i}^{2}\right\rangle }{i\omega\left\langle \Delta\hat{R_{i}^{2}}(\omega)\right\rangle }-1\right]\label{eq:12}
\end{equation}

$G^{*}(\omega)$ is the frequency dependent dynamic complex modulus,
the real part (elastic modulus) of which represents the amount of
energy stored and the imaginary part (viscous modulus) represents
dissipation of energy. $k_{i}$ is the stiffness of the i-th trap
in which the probe is confined, $a_{i}$ is the radius of the probe.
$\left\langle \Delta\hat{R_{i}^{2}}(\omega)\right\rangle $ is the
Fourier transform of the time dependent MSD of the thermal position
fluctuation of the probe particle. $\left\langle R_{i}^{2}\right\rangle $
is the time independent variance.

For a single trapped particle in a viscous medium, the auto-correlation
is given by

\begin{equation}
\left\langle R_{i}(\tau)R_{i}(0)\right\rangle =\frac{K_{B}T}{k_{i}}\exp(-\omega_{c}\tau)\label{eq:13}
\end{equation}

where, $\omega_{c}=\frac{k_{i}}{6\pi\eta a_{i}}$. The time-dependent
mean-square displacement is 

\begin{eqnarray}
\left\langle \Delta R_{i}^{2}(\tau)\right\rangle  & = & 2\left[\left\langle R_{i}^{2}(0)\right\rangle -\left\langle R_{i}(\tau)R_{i}(0)\right\rangle \right]\label{eq:17_1}\\
 & = & \frac{2K_{B}T}{k_{i}}\left[1-\exp\left(-\omega_{c}\tau\right)\right]\label{eq:14}
\end{eqnarray}

Now,

\begin{eqnarray*}
G^{*}(\omega) & = & G(s)|_{s=i\omega}\\
 & = & \frac{k_{i}}{6\pi a_{i}}\left[\frac{2\left\langle R_{i}^{2}\right\rangle }{s\left\langle \Delta\hat{R_{i}^{2}}(s)\right\rangle }-1\right]\Biggl|_{s=i\omega}\\
 & = & i\eta\omega
\end{eqnarray*}

Hence, the surrounding medium accessed by the single trapped particle
is purely dissipative in nature. Now, if we go through similar process
assuming one of the pair of trapped particles as probe then

\begin{widetext}

\begin{alignat}{1}
G^{*}(\omega)= & \frac{k_{i}}{6\pi a_{i}}\Bigg[\omega^{2}\Biggl\{\frac{(A\beta_{-}+B\beta_{+})(A\beta_{+}+B\beta_{-})}{\omega^{2}(A\beta_{+}+B\beta_{-})^{2}+\beta_{+}^{2}\beta_{-}^{2}(A+B)^{2}}-\frac{\beta_{+}\beta_{-}(A+B)^{2}}{\omega^{2}(A\beta_{+}+B\beta_{-})^{2}+\beta_{+}^{2}\beta_{-}^{2}(A+B)^{2}}\Biggr\}\nonumber \\
 & +i\omega\Biggl\{\frac{(A+B)(A\beta_{-}+B\beta_{+})\beta_{+}\beta_{-}}{\omega^{2}(A\beta_{+}+B\beta_{-})^{2}+\beta_{+}^{2}\beta_{-}^{2}(A+B)^{2}}+\frac{\omega^{2}(A+B)(A\beta_{+}+B\beta_{-})}{\omega^{2}(A\beta_{+}+B\beta_{-})^{2}+\beta_{+}^{2}\beta_{-}^{2}(A+B)^{2}}\Biggr\}\Biggr]\label{eq:15}
\end{alignat}

\end{widetext}

where A and B are the amplitudes of the exponentials of the auto-correlation
function of the probe. 

\section{Experimental setup}

\begin{figure}[!tbp]
\label{fig1}\includegraphics[scale=0.4]{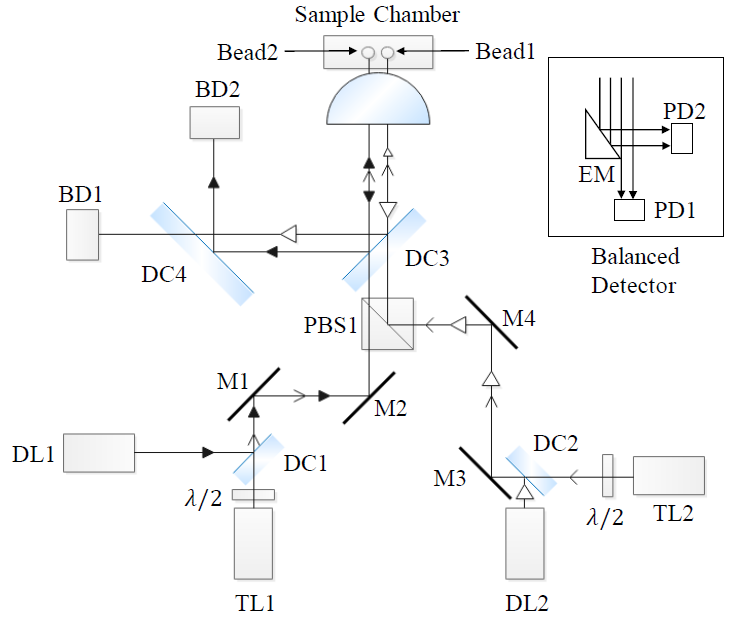}

\caption{Schematic of the experimental setup. $\lambda/2$: half-wave plate,
M: plane mirror, DC: dichroic, TL: trapping laser, DL: detection laser,
PBS: polarizing beam splitter, BD: balance detector, EM: edge mirror,
PD: photo-diode (Thorlabs PDA100A-EC).}

\label{fig:1}
\end{figure}
To check the validity of the above-described theory we set up (\ref{fig1})
a dual-beam optical tweezers by focusing two independently generated
orthogonally polarized laser beams from two diode lasers (TL1 and
Tl2) of wavelength $\lambda=1064$ nm, using a high NA immersion-oil
microscope objective (Zeiss PlanApo,$100\times1.4$). Two $\lambda/2$
plates, in front of the lasers control the polarization angle of these
two laser beams. After passing through $\lambda/2$ plates these two
beams encounter mirror pairs M1, M2 and M3, M4, respectively, as two
beam steerers. Then they get coupled into a polarizing beam splitter
(PBS1). For detection, we use two lasers of wavelength $\lambda=671$
nm (DL1) and $\lambda=780$ nm (DL2) which we couple to the trapping
lasers using two dichroic mirrors (DC1 and DC2) before the beam steering mirrors. We image two trapped beads and measure their displacements by back-focal-plane-interferometry,
while we use white light for imaging. After collecting the total back-scattered
light from the microscope, we separate the components from the two particles using another dichroic (DC4) and then direct these towards two balanced detection systems BD1 and BD2, developed using photodiode pairs. The cartoon representation of one such balanced detector is shown in the inset of Fig.~\ref{fig:1}. The voltage-amplitude calibration of our detection
system reveals that we can resolve motion of around 5 nm with an SNR
of 2. We prepared a very low volume fraction sample ($\phi\approx0.01$)
with 3 $\mu$m diameter polystyrene latex beads in 1 M NaCl-water
solution for avoiding surface charges. We loaded the sample in a chamber
of area $20\times10$ mm and height 0.2 mm and two spherical polystyrene
beads (Sigma LB-30) of mean size 3 $\mu$m each were trapped in two
calibrated optical traps which we kept separated initially by a distance
of $5\pm0.1$ micron from each other, and at a distance of 30 $\mu$m
from the nearest wall. Then the separation and laser powers were varied
to perform our experiment. According to a reported experimental work
\cite{stilgoe2011phase}, this distance is large enough to avoid optical
cross-talk and the effects from surface charges. In order to ensure
that the trapping beams do not influence each other, we measured the
Brownian motion of one when the other is switched on (in the absence
of a particle), and checked that there were no changes in the properties
of the Brownian fluctuations. We normalized each time series representing
the position fluctuation of each particle by the sum intensity measured
by the corresponding photodiode pair to account for the laser power
fluctuations. Then we collected it, typically over 10 second and at sampling
frequency 10kHz using a data acquisition card (NI USB-6356), which
was coupled to a computer. we recorded two time series
corresponding to two trapped particles simultaneously. We check that there is less than one percent cross-talk of signals in the two detectors due to
leakage through the dichroic DC4. 

\bibliographystyle{apsrev4-1}
%